\documentclass[aps,prl,reprint,superscriptaddress]{revtex4-2}
\usepackage{graphicx}
\usepackage{amsmath}
\usepackage{amssymb}
\usepackage{xcolor}

\begin{document}

\title{Competing Auger-Like and Intraband Excitations Drive Anomalous Modulations in Indium Tin Oxide}

\author{Anthony C. Harwood}
\email{a.harwood22@imperial.ac.uk}
\affiliation{Blackett Laboratory, Department of Physics, Imperial College London, London SW7 2AZ, UK}

\author{Sean Z. J. Lim}
\affiliation{Blackett Laboratory, Department of Physics, Imperial College London, London SW7 2AZ, UK}

\author{T. V. Raziman}
\affiliation{Blackett Laboratory, Department of Physics, Imperial College London, London SW7 2AZ, UK}
\affiliation{Department of Mathematics, Imperial College London, London, UK}

\author{Yan Li}
\affiliation{Blackett Laboratory, Department of Physics, Imperial College London, London SW7 2AZ, UK}

\author{Joseph Stones}
\affiliation{Blackett Laboratory, Department of Physics, Imperial College London, London SW7 2AZ, UK}

\author{John W. G. Tisch}
\affiliation{Blackett Laboratory, Department of Physics, Imperial College London, London SW7 2AZ, UK}

\author{Simon A. R. Horsley}
\affiliation{School of Physics and Astronomy, University of Exeter, Exeter EX4 4QL, UK}

\author{Stefano Vezzoli}
\affiliation{Blackett Laboratory, Department of Physics, Imperial College London, London SW7 2AZ, UK}

\author{John B. Pendry}
\affiliation{Blackett Laboratory, Department of Physics, Imperial College London, London SW7 2AZ, UK}

\author{Riccardo Sapienza}
\affiliation{Blackett Laboratory, Department of Physics, Imperial College London, London SW7 2AZ, UK}

\date{\today}

\begin{abstract}
\textbf{Abstract:}~Transparent conducting oxides near their epsilon-near-zero frequency exhibit near-unity ultrafast modulations of the refractive index which have enabled the field of time-varying metamaterials, yet the underlying carrier dynamics at high driving fluences remain poorly understood. Here, we report ultrafast modulations in the reflectivity and transmissivity of indium tin oxide, and a non-monotonic oscillatory behavior. This is especially evident in the time evolution of the complex Fresnel coefficients retrieved directly from pump-probe spectrograms using an optical gating technique, GRUMPY FROG. The dynamics of the retrieved plasma frequency and damping coefficient are well captured by an extended two-temperature model incorporating a competing nonlinear interband process: at high fluences, Auger-type scattering of hot conduction electrons promotes valence band carriers, increasing the plasma frequency while accelerating hot-electron cooling and raising the damping coefficient. These results clarify the origin of anomalous high-fluence dynamics in indium tin oxide and identify a fluence-tuneable modulation dynamic with direct implications for ultrafast refractive index engineering in time-varying photonic devices and optical switching.
\end{abstract}

\maketitle
\section{Introduction}

Temporally structured media promise a new paradigm for the control of optical signals, enabling non-reciprocal devices, spectral manipulation, beam steering and amplification via near-unity modulations of material properties at optical frequencies \cite{galiffiPhotonicsTimevaryingMedia2022, enghetaFourdimensionalOpticsUsing2023}. Experimental progress towards realising these effects has been dominated by the excitation of transparent conducting oxide (TCO) thin films with ultrashort laser pulses near their epsilon-near-zero (ENZ) wavelength, where vanishing real permittivity and high optical damage thresholds have been leveraged to provide near-unity, sub-picosecond modulations of refractive index \cite{alamLargeOpticalNonlinearity2016,caspaniEnhancedNonlinearRefractive2016, kinseyEpsilonnearzeroAldopedZnO2015} alongside a large, non-perturbative third-order nonlinearity \cite{caprettiEnhancedThirdharmonicGeneration2015, reshefPerturbativeDescriptionNonlinear2017}. By utilising structured temporal and spatiotemporal modulations, TCOs have underpinned novel demonstrations of optical switching \cite{guoLargeOpticalNonlinearity2016, clericiControllingHybridNonlinearities2017}, temporal refraction \cite{lustigTimerefractionOpticsSingle2023, zhouBroadbandFrequencyTranslation2020, liuTunableDopplerShift2021}, diffraction from synthetically moving modulations \cite{vaddiParametricAmplificationOptical2026, harwoodSpacetimeOpticalDiffraction2025}, coherent wave control \cite{galiffiOpticalCoherentPerfect2026}, and ultrafast polarisation manipulation \cite{jaffrayAllopticalPolarizationControl2026}.

The ultrafast photo-induced response of TCOs, such as indium tin oxide (ITO), has primarily been studied through pump-probe spectroscopy. In these experiments, pump photons with energies below, or above, the direct bandgap (for ITO this is 3.9~eV) modulate the material either by changing effective mass of conduction electrons ($m^*$)~\cite{alamLargeOpticalNonlinearity2016} or the conduction band electron density ($n_c$)~\cite{kinseyEpsilonnearzeroAldopedZnO2015, clericiControllingHybridNonlinearities2017}, respectively. However, due to the accessibility of high-powered pulsed lasers in the NIR and the difficulties involved with using UV pump pulses, investigations have primarily focussed on intraband modulations. In such experiments, up until moderate pump fluences, the material response has been well described by a two-temperature model combined with the Drude model, in which nearly instantaneous intraband absorption promotes conduction electrons to higher-energy states within the non-parabolic band. This process keeps the carrier density ($n_c$) constant but increases the effective mass of conduction band electrons ($m^*$), reducing the plasma frequency $\omega_\mathrm{p} = \sqrt{n_c e^2/m^*\varepsilon_0}$, before recovery via a slower, approximately 300~fs, electron-phonon mediated process \cite{alamLargeOpticalNonlinearity2016}. These models assume an electron temperature-dependent shift in $\omega_\mathrm{p}$, and have been used to great effect across numerous experimental works \cite{bohnAllopticalSwitchingEpsilonnearzero2021, bohnSpatiotemporalRefractionLight2021b, guoLargeOpticalNonlinearity2016, secondoAbsorptiveLossBand2020, gurungControlUltrafastHot2025}, notably capturing the saturation of the $\omega_\mathrm{p}$ shift with increasing pump energy \cite{khurginNonperturbativeNonlinearitiesPerhaps2024, liUltrafastSwicthingOptical}. However, for high fluences, the two-temperature model breaks down. 

Several aspects of ITO's behaviour under intense pumping resist this standard description. Standard two-temperature modelling predicts extreme electron temperatures that should cause catastrophic sample damage, yet ITO and related TCOs exhibit anomalously high damage thresholds despite large energy absorption \cite{unElectronicBasedModelOptical2023, wangRoleHotElectron2020}. Moreover, recent experiments at high pump fluences have revealed both ultrafast decay dynamics \cite{lustigTimerefractionOpticsSingle2023, liUltrafastSwicthingOptical} and oscillations~\cite{ozluNonlinearityReversalEpsilonNearZero2026, jaffraySinglepumpHybridNonlinearities2026}, first reported by Segal \textit{et al.}~\cite{segalComplexTemporalDynamics2025}, whose origins remain actively debated \cite{narimanovUltrafastOpticalModulation2025, choiPathwayOpticalCycleDynamic2025, unQuantumopticalTheoryFewfemtosecond2026}. Understanding these behaviours requires independent, time-resolved access to both $\omega_{p}(t)$ and $\gamma(t)$. While $\omega_{p}$ dynamics have been characterised extensively \cite{alamLargeOpticalNonlinearity2016, caspaniEnhancedNonlinearRefractive2016}, $\gamma(t)$ is more rarely considered \cite{gurungControlUltrafastHot2025, wangRoleHotElectron2020}, despite theoretical evidence that it plays a significant role \cite{unOpticalNonlinearityTransparent2025}, and neither has been resolved in the high-fluence regime. In particular, extracting $\omega_p(t)$ resolves the interplay between $m^*$ and $n_c$, making it sensitive to both intraband and interband processes that inject free carriers into the conduction band. While traditionally observed via above-bandgap, ultraviolet pumping \cite{guizzardiLargeScaleIndium2020, clericiControllingHybridNonlinearities2017, kinseyEpsilonnearzeroAldopedZnO2015}, Li and colleagues recently attributed the complex modulation dynamics of a 310 nm ITO film under below-bandgap pumping to a competing interband process driven by extreme intraband heating \cite{liUltrafastSwicthingOptical}, challenging the prevailing picture and motivating a detailed investigation.

Here we perform pump-probe spectroscopy on a 310~nm sample of ITO and observe ultrafast modulations in the transmissivity and reflectivity, induced at pump fluences greater than 55~mJ/cm$^{2}$, which cannot be explained by the standard two-temperature model. We develop a novel pump-probe retrieval technique called Generalised Retrieval of Ultrafast Modulations from Pump-probe Yields Frequency Resolved Optical Gating (GRUMPY FROG) to extract ultrafast modulations of the Fresnel coefficients. We reconstruct the time-dependent $\omega_\mathrm{p}$ and $\gamma$ via the inverse transfer matrix method and observe competing intra-/interband contributions, sub-100~fs relaxation of both the transmissivity and reflectivity, and large $\gamma$ changes. We model these dynamics with a two-temperature model that we combine with an Auger interband process, which quantitatively reproduces all observed $\omega_\mathrm{p}$ features across the full fluence range. Moreover, the presence of Auger excitation at high fluences acts as a heat sink for conduction electrons, limiting their temperature far below two-temperature model predictions and provides a mechanism for fluence-dependent ultrafast refractive index engineering near the ENZ condition.

\section{Results}

\begin{figure}[t]
\centering
\includegraphics[width=\linewidth]{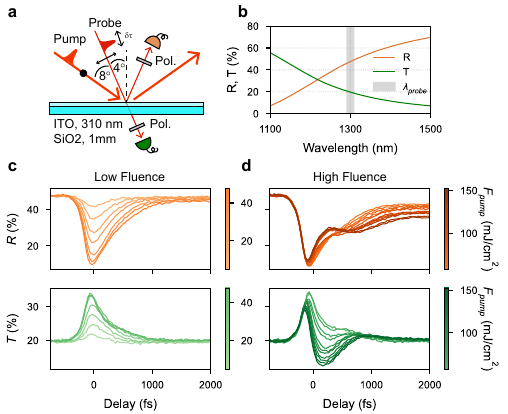}
\caption{\textbf{Pump-probe spectroscopy of a sub-wavelength ITO film.}
(a)~Schematic of the degenerate pump-probe setup. A 1300~nm pump pulse excites the ITO film at near-normal incidence while a cross-polarised probe pulse, delayed by $\delta\tau$, is reflected or transmitted and spectrally resolved.
(b)~Reflectivity $R$ and transmissivity $T$ of the 310~nm ITO film as a function of wavelength, with the probe bandwidth at 1300~nm indicated (blue dashed line).
(c,\,d)~Pump-probe traces for reflection (orange) and transmission (green) as functions of pump-probe delay, shown across pump fluences from 4.6 to 153~mJ/cm$^{2}$. Below 55~mJ/cm$^{2}$ (c), the modulation follows the typical two-timescale intraband response. Above this fluence (d), fast decay dynamics and oscillatory behaviour emerge.}
\label{fig:setup}
\end{figure}

We use an angularly separable, degenerate pump-probe configuration (Fig.~\ref{fig:setup}a) to investigate below-bandgap optical modulation of a 310~nm ITO film deposited on a glass substrate using 136~fs pulses (see SI for measured SHG-FROG) centred at 1300~nm (230~THz), near the ENZ wavelength of the sample. The pump and probe beams are incident at near-normal incidence (probe at $4^\circ$ and pump at $8^\circ$), cross-polarised with s- and p-polarisation respectively, with an analyser at the output isolating the reflected probe from pump scattering and two-beam coupling. Pump fluence is varied from 4.6 to 153~mJ/cm$^{2}$, with the laser operating at a repetition rate of 20~Hz to allow sufficient thermal recovery between pulses and avoid sample damage through accumulated heating at the highest fluences~\cite{herbstUltrafastNonlinearDynamics2026}. Throughout all datasets, the probe fluence is kept to approximately 0.1~mJ/cm$^{2}$ to avoid unwanted probe-induced nonlinearities. Pump-probe signals are measured in both reflection and transmission, with the sample demonstrating high dispersion in both configurations (Fig.~\ref{fig:setup}b). 

In the regime of low fluence, below 55~mJ/cm$^{2}$ (Fig.~\ref{fig:setup}c), absorption of the pump pulse heats and promotes electrons within the conduction band, which immediately thermalise via electron-electron scattering before equilibrating via electron-phonon scattering, a process well matched by a two-temperature model \cite{unElectronicBasedModelOptical2023}. Due to the material's non-parabolic band structure, redistribution of conduction-band electrons to higher energies increases their effective mass, which consequently shifts $\omega_\mathrm{p}$ towards lower frequencies. For a probe centred at 1300~nm, this shift in $\omega_p$ is measured as an increase in transmissivity and a decrease in reflectivity, since the dispersion curves shift to longer wavelength. The rise time is on the order of the pump pulse duration, with a decay of approximately 300~fs \cite{alamLargeOpticalNonlinearity2016}. As the top end of this low-fluence range is reached, the magnitude of the modulation begins to saturate and the rise time decreases, a feature that has been well-documented \cite{tiroleSaturableTimeVaryingMirror2022,caspaniEnhancedNonlinearRefractive2016} and attributed to the depopulation of electrons from below the Fermi level, whose relative change in effective mass is greatest \cite{liUltrafastSwicthingOptical}.

However, for fluences greater than 55~mJ/cm$^{2}$ (Fig.~\ref{fig:setup}d), the observed transmission and reflection dynamics depart from the two-timescale intraband-only model. Rather than evolving with the shifting of the static dispersion curve as in the low-fluence regime, a secondary ultrafast process emerges that works antagonistically to the intraband process while impacting the reflection and transmission asymmetrically. This asymmetry is an indication that $\gamma$, not just $\omega_p$, is being significantly modulated at high fluences. In transmission, at fluences of approximately 70 mJ/cm$^{2}$, the modulation dynamics and fast decay resemble the traces reported by Lustig et al. \cite{lustigTimerefractionOpticsSingle2023} and Li et al.~\cite{liUltrafastSwicthingOptical}. Here the absorption of the 1300~nm pump is much greater than that of the 800~nm pump used in the aforementioned experiments~\cite{lustigTimerefractionOpticsSingle2023,liUltrafastSwicthingOptical}. As the fluence further increases, the speed of the decay mechanism increases before transiently plunging below half of its static value at our highest fluence of 153~mJ/cm$^{2}$. Prior works have also observed oscillatory transients in TCOs, due to crossing of the sample's plasmon resonance \cite{bohnAllopticalSwitchingEpsilonnearzero2021}, antagonistic intraband and interband modulations \cite{clericiControllingHybridNonlinearities2017}, or coherent contributions from two-beam coupling \cite{paulTwobeamCouplingHot2021, vaddiParametricAmplificationOptical2026}. By operating far from the plasmon resonance, using pump photons with energies less than the bandgap, and working with cross-polarised beams, we isolate our measurements from these phenomena. 

Here we introduce GRUMPY FROG (Generalised Retrieval of Ultrafast Modulations from Pump-probe Yields FROG), a sub-variety of the FROG pulse retrieval menagerie \cite{trebinoMeasuringUltrashortLaser1997} that extracts the transient Fresnel coefficient directly from the acquired pump-probe spectrogram $I(\omega, \tau)$ which we can convert into $\omega_\mathrm{p}(t)$ and $\gamma(t)$. FROG techniques use iterative algorithms to retrieve the full complex electric field of ultrashort pulses, both amplitude and phase, from a measured spectrogram formed by gating the unknown pulse with another through a nonlinear optical interaction \cite{delongUltrashortpulseMeasurementUsing1995, delongImprovedUltrashortPulseretrieval1994, jafari100ReliableAlgorithm2019, wongSingleshotMeasurementComplete2014}. For nonlinear phenomena such as sum-frequency generation, the polarisation response is the product of the electric field of the unknown pulse $E(t)$ with a delayed copy of itself or with a known gate $G(t-\tau)$,
\begin{equation}
  I(\omega,\tau) = \left|\int_{-\infty}^{\infty} E(t)\cdot G(t-\tau)\,e^{-i\omega t}\,\mathrm{d}t\right|^2.
  \label{eq:frog}
\end{equation}

In GRUMPY FROG the probe field $E(t)$ is gated by an effective time-varying, complex and adiabatic Fresnel coefficient, for example reflection, defined as $r(t) = |r|\,e^{i\varphi(t)}$,
\begin{equation}
  I(\omega,\tau) = \left|\int_{-\infty}^{\infty} E(t)\cdot r(t-\tau)\,e^{-i\omega t}\,\mathrm{d}t\right|^2.
  \label{eq:grumpyfrog}
\end{equation}
Here, reflection serves as the mediating instantaneous process provided the film can be approximated as non-dispersive across the probe bandwidth, enabling $r(t)$ to be treated as a scalar gate rather than a frequency-dependent operator. We verify this assumption through operator-based time-varying dispersive simulations \cite{horsleyEigenpulsesDispersiveTimeVarying2023}, with full details given in the supplementary information. The acquired pump-probe spectrograms are therefore described by Eq.~(\ref{eq:grumpyfrog}), enabling retrieval of both the complex reflectivity modulation induced by the pump and the incident probe pulse via well-known algorithms. A more detailed discussion of GRUMPY FROG is given in the supplementary information.

\begin{figure}[t]
\centering
\includegraphics[width=\linewidth]{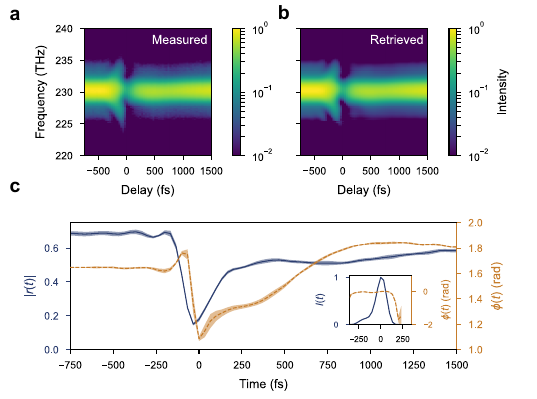}
\caption{\textbf{GRUMPY FROG retrieval of the complex Fresnel reflection coefficient.}
(a)~Measured and (b)~retrieved pump-probe spectrogram at a pump fluence of 133~mJ/cm$^{2}$. The input probe spectrum is naturally embedded at large negative delays and the spectrogram develops pronounced asymmetric structure about zero delay as the pump modulates the sample.
(c)~Retrieved complex reflection coefficient $r(t)$ as a function of pump-probe delay, normalised using independently measured ellipsometry, showing the amplitude $|r(t)|$ (blue, left axis) and phase $\varphi(t)$ (orange, right axis) with shaded uncertainties measured via bootstrapping~\cite{wangDeterminingErrorBars2003}. Both observables exhibit a fast initial modulation followed by a sharp recovery and oscillations at positive delays far beyond the duration of the pump. The simultaneously retrieved probe pulse (inset) is consistent with independent SHG-FROG measurements.}
\label{fig:grumpyfrog}
\end{figure}

It is important to note that non-instantaneous nonlinearities have been used to mediate FROG measurements \cite{delongUltrashortpulseMeasurementUsing1995} and ultrafast dynamics of media have been inferred by measuring the change a pulse scattered by a dynamic medium \cite{jonesMeasuringUltrashortUltraviolet2021, rodriguezCoherentUltrafastMIFROG2001, fruhlingSpectralHolographyTimeRefraction2025}. Additionally, sub-picosecond modulations of media have been used as a broadband gate for measuring the space-time intensity profile of optical pulses\cite{ballSpaceTimeKnifeEdgeEpsilonNearZero2024a}, as well as measuring the complex envelopes of optical pulses \cite{leblancPhasematchingfreePulseRetrieval2019, cilentoUltrafastInsulatortometalPhase2010}. However, in the latter experiments, the extracted gate is a purely temporal approximation $r(t)$ of the full time-varying dispersive response $r(\omega, t)$, averaging over the frequency dependence. In contrast, we use GRUMPY FROG as a technique for characterising the ultrafast dynamics of media, not optical pulses, by retrieving the modulation from spectrograms acquired under non-dispersive conditions.

\begin{figure}[t]
\centering
\includegraphics[width=\linewidth]{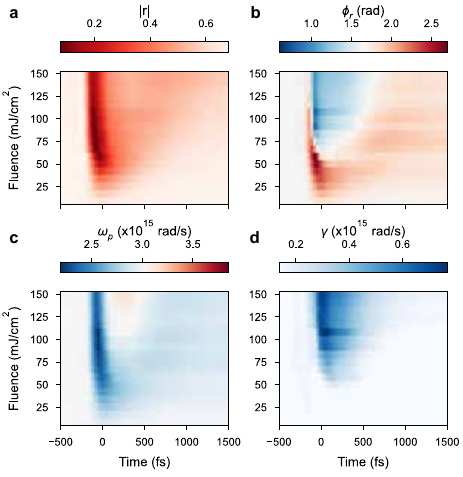}
\caption{\textbf{Fluence-dependent evolution of the retrieved complex reflectivity and extracted Drude parameters.}
(a,\,b)~Two-dimensional colourmaps of the retrieved reflectivity amplitude $|r(t)|$ and phase $\varphi(t)$ as functions of pump-probe delay and pump fluence. At low fluences both observables show a smooth, single-signed modulation with a fast rise and slower recovery consistent with intraband heating and electron-phonon coupling. Above approximately 55~mJ/cm$^{2}$ the amplitude minimum saturates, while the phase develops a sign reversal on the trailing edge, indicating a competing process pushing the film to the other side of ENZ.
(c,\,d)~Corresponding colourmaps of the effective plasma frequency $\omega_\mathrm{p}(t)$ and Drude damping $\gamma(t)$, extracted from $r(t)$ via the inverse transfer matrix method. At low fluences $\omega_\mathrm{p}$ shows the expected intraband decrease and recovery. At high fluences a delayed positive contribution drives $\omega_\mathrm{p}$ transiently above its equilibrium value, while $\gamma(t)$ develops modulations with peaks 700\% above the equilibrium value.}
\label{fig:drude_maps}
\end{figure}

Fig.~\ref{fig:grumpyfrog}a,b shows the measured and retrieved pump-probe spectrograms in reflection at a pump fluence of 133~mJ/cm$^{2}$, with the root-mean-squared error between the two traces being 0.3\% (see SI for more details on the retrieval accuracy and validation). The transient complex reflectivity $r(t)$ in Fig.~\ref{fig:grumpyfrog}c is then obtained by scaling the normalised retrieved trace by the static Fresnel coefficient measured independently via ellipsometry. By repeating this process across all pump fluences, we extract the amplitude $|r(t)|$ and phase $\varphi(t)$ as functions of time and pump fluence, displayed in Figs.~\ref{fig:drude_maps}a and b. Although we retrieve both reflection and transmission coefficients (see SI), we focus here on reflection, as the signal is dominated by the air-ITO interface and free from nonlinear effects such as cross-phase modulation that may arise in the 1~mm SiO$_2$ substrate at high fluences \cite{bresciRemovalCrossphaseModulation2021}. At low fluences both amplitude and phase show a smooth, single-signed modulation centred near zero delay with a $\sim$300~fs recovery consistent with electron-phonon coupling. As fluence increases, the amplitude minimum deepens and sharpens, approaching zero at the highest fluences. Above approximately 55~mJ/cm$^{2}$ the amplitude develops a clear asymmetry, partially recovering before the expected electron-phonon timescale, while the phase undergoes a sign reversal at approximately $-100$~fs, producing a positive lobe on the trailing edge of the modulation.

Describing the transient permittivity within a Drude model, we invert the complex reflectivity via the transfer matrix method to extract $\omega_\mathrm{p}(t)$ and $\gamma(t)$ directly, shown in Figs.~\ref{fig:drude_maps}c and d. These extracted parameters are qualitatively in agreement with inversions from the complex transmission data, in spite of additional phase due to cross-phase modulation in the substrate. The TMM inversion treats the film as uniformly modulated, such that the extracted parameters represent effective values of the spatially  stratified film (we validate this approximation in the supplementary information). At low fluences $\gamma$ shows minimal variation and $\omega_\mathrm{p}$ exhibits a smooth modulation saturating near $-20\%$, in quantitative agreement with Li et al. \cite{liUltrafastSwicthingOptical}. Above 55~mJ/cm$^{2}$ both parameters depart from this behaviour: the $\omega_\mathrm{p}$ minimum shallows and reverses sign as a secondary process drives the plasma frequency above its equilibrium value by over 3\%, while $\gamma$ develops transient peaks exceeding 600\% of its unperturbed value. The coincidence of both thresholds points to a common underlying mechanism.

The retrieved $\omega_\mathrm{p}$ maps in Fig.~\ref{fig:drude_maps}c reveal a strong divergence from the two-temperature model for pump fluences greater than 55~mJ/cm$^{2}$, where the $\omega_\mathrm{p}$ minimum shallows and reverses, with a delayed positive excursion that drives $\omega_\mathrm{p}$ above its equilibrium value by over 4\%. A conduction-band-only model predicts only a monotonic decrease in $\omega_\mathrm{p}$ as pump energy is deposited. We postulate that the observed sign reversal originates through Auger excitation: heating conduction electrons raises them into states of higher effective mass, reducing $\omega_\mathrm{p}$, but further heating delivers sufficient energy for valence electrons to be promoted into the conduction band, increasing electron density and driving $\omega_\mathrm{p}$ above its equilibrium value. 

\begin{figure}[t]
    \centering
    \includegraphics[width=\linewidth]{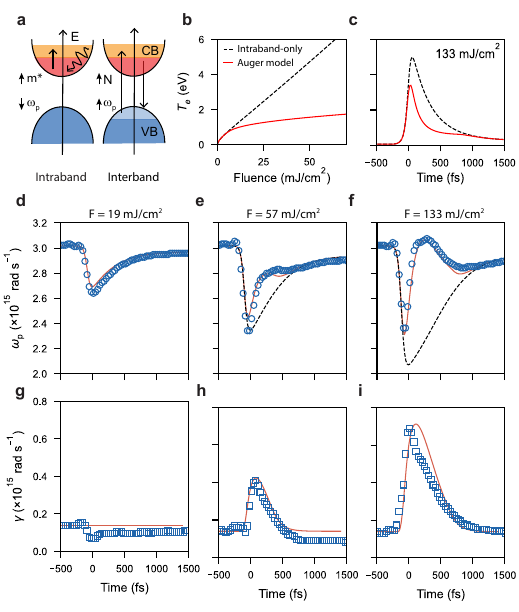}
    \caption{\textbf{Competing intraband and interband carrier dynamics in ITO.}
    (a)~Intraband absorption increases the effective mass $m^*$ of conduction-band (CB) electrons and reduces $\omega_\mathrm{p}$. At high fluences, Auger excitation promotes carriers from the valence band (VB) into the conduction band, increasing free-carrier density $N$ and driving $\omega_\mathrm{p}$ upward.
    (b) Equilibrium conduction-electron temperature versus pump fluence calculated using the intraband-only model (dashed line) and the combined intraband and Auger model (full line). In each case perfect thermal equilibrium is assumed. (c) as in (b) but in this case the delayed onset of Auger electrons is taken into account. Note the short-lived spike in temperature in the full line, rapidly damped by Auger processes.
    (d--f)~Retrieved $\omega_\mathrm{p}(t)$ (circles) compared to model predictions with (full) and without (dashed) interband contributions at 19, 57 and 133~mJ/cm$^{2}$.(g--i)~Retrieved $\gamma(t)$ (circles) compared to the calculated density of holes in the valence band (full).}
    \label{fig:model}
\end{figure}

With involvement of valence electrons, the assumption of thermal equilibrium among electrons breaks down. Were the system to remain in thermal equilibrium throughout, each electron temperature would be experienced twice, once during heating and once during cooling; thus, the $\omega_\mathrm{p}$ trace would exhibit two minima of identical depth. The asymmetry of the $\omega_\mathrm{p}$ minima at high fluences, clearly visible in Fig.~\ref{fig:drude_maps}c, rules this out. In our model, just as there is an energy imbalance between conduction electrons and phonons, there is also an imbalance between conduction and valence electrons. Auger excitation is slower than intraband thermalisation: the interband scattering rate does not permit immediate equilibration between conduction and valence bands, and a delay relative to the pump excitation is therefore expected~\cite{ziminPlasmascopyUltrafastHot2025,bridaUltrafastCollinearScattering2013a}. We measure the peak of the interband contribution at a delay of 120~fs, which is a key characteristic of Auger excitation~\cite{ziminPlasmascopyUltrafastHot2025, bridaUltrafastCollinearScattering2013a}, and a central feature of the dynamics that the model must reproduce. Two-photon absorption, sometimes proposed as an alternative interband mechanism, operates only during the pump pulse and is ruled out by this delay, which exceeds the pump duration. Fig.~\ref{fig:model}b illustrates schematically how the competition between intraband and delayed interband transitions produces the characteristic dynamics observed in Fig.~\ref{fig:drude_maps}c: intraband heating drives $\omega_\mathrm{p}$ downward during the pump, before the delayed increase in carrier density partially reverses this decrease, lifting $\omega_\mathrm{p}$ transiently above equilibrium. Interband transitions also offer an efficient route to relax the energy of the hot electron gas. The resulting equilibrium temperatures are shown in Fig.~\ref{fig:model}c as a function of deposited energy. Without interband promotion (dashed line), the temperature rises without bound, exceeding 5~eV at the maximum fluence. Including Auger excitation (solid line), the large valence-band density of states acts as a heat reservoir of enormous capacity, capping the electron temperature below 1.5~eV, offering an explanation for why the extreme fluences used in TCO-based ultrafast experiments do not cause sample damage.

We describe the conduction-band dispersion using a modified Kane model \cite{kaneBandStructureIndium1957} with parameters from \cite{mryasovElectronicBandStructure2001, medvedevaTuningPropertiesComplex2010}, giving the effective mass as a function of electron energy below the Fermi level. Above the Fermi level the band flattens and $m^*$ is approximately constant; this high-energy effective mass governs the saturation of the $\omega_\mathrm{p}$ shift at large pump fluences \cite{liUltrafastSwicthingOptical}, and we adjust it by a factor of 0.75 relative to the Kane extrapolation to match the experiment. The valence bands are known to have a very high density of states immediately below the band edge, but in the absence of quantitative data we treat the density of valence states as a parameter to fit the weight of the Auger contribution. Both parameters, once fitted at the highest pump fluence, remain unchanged throughout further fitting.

As in the usual two-temperature model, our model assumes thermal equilibrium among conduction electrons. However, due to the high energies of conduction electrons at larger pump power, we allow the number of conduction electrons to increase via the process of Auger excitation, whereby conduction and valence electrons collide, exciting valence electrons across the band gap into the conduction band, leaving behind holes in the valence band. This leads to a decrease in energy, but an increase in density of the conduction electrons and over time this increased carrier density decays back to its equilibrium value via Auger recombination.

We model this process with a simple linear rate equation for the density of conduction electrons.  The excitation rate is taken as proportional to the difference between $n_0$, the conduction electron density assuming perfect equilibrium with the valence band at temperature $T$, and the actual non-equilibrium conduction electron density $n_c$ at time $t$ and temperature $T$,
\begin{equation}
  \dot{n}_c = \eta\left[n_0 - n_c\right],
  \label{eq:auger}
\end{equation}
where $\eta$ is the Auger excitation rate, to which we assign the value 1/120~fs$^{-1}$.  The motivation for Eq. (\ref{eq:auger}) can be understood as follows:  when the thermal energy of the conduction electrons becomes comparable to the band gap energy of ITO, the right hand side of (\ref{eq:auger}) will be positive ($n_0>n_c$) and significantly different from zero, leading to an increase in the density of conduction electrons, at a rate $\eta$.  As described below, this process removes energy from (cools) the conduction electrons until the right hand side of (\ref{eq:auger}) is negative, at which point the system begins its decay back to equilibrium.  Given the simplicity of Eq. (\ref{eq:auger}) we can solve it analytically, giving $n_c(t)=\eta\int_{-\infty}^{t}{\rm d}t'\,{\rm e}^{-\eta(t-t')}n_0(t')$.  Further details are given in the Supplementary Material.

In the two-temperature model, energy is deposited by the pump to the conduction electrons, and drains slowly to the phonons. We thus specify that, without the excitation of valence electrons, the conduction electrons have the time dependent energy density
\begin{equation}
  U_0(t) = \frac{1}{2}U_{\max}\left[e^{-\beta t} + U_{\infty}/U_{\max}\right]\left[\tanh(\alpha t) + 1\right],
  \label{eq:ttm}
\end{equation}
where $\alpha$ = 1/50~fs$^{-1}$ is the inverse rise time of energy as determined by the duration of the pump.  Decay by phonon emission is determined by $\beta$ = 1/300~fs$^{-1}$ \cite{alamLargeOpticalNonlinearity2016}, which we fix to its low-fluence value to focus on the fitting of the interband contribution. It has been reported \cite{guoLargeOpticalNonlinearity2016, liUltrafastSwicthingOptical} that, in addition to the long phonon decay described by $\beta^{-1}$, there is an even longer timescale, perhaps as long as milliseconds, during which residual energy in the system keeps $\omega_\mathrm{p}$ depressed. The mechanism for this energy retention is not understood; we account for it through a residual energy $U_{\infty}$ adjusted to match the experiment at long delays.

In contrast to the two-temperature model, the increase/decrease in conduction electron density via (\ref{eq:auger}) is accompanied by a decrease/increase in conduction electron energy.  We model this through writing the rate of change of the conduction electron energy $U_c$ with an additional term compared to (\ref{eq:ttm})
\begin{equation}
  \dot{U}_c(t) = \dot{U}_0(t) - E_g\dot{n}_c(t).
  \label{eq:energy}
\end{equation}
The valence electrons have a massive density of states at the band edge, allowing us to assume that most excitations are from the edge and therefore require an input of the band gap energy, $E_{g}$ = 3.9 eV, for each electron excited.

From equations~(\ref{eq:auger}--\ref{eq:energy}) we obtain both the number density $n_c(t)$ and energy density $U_{c}(t)$ of the conduction electrons. Assuming a Fermi--Dirac distribution applies to the conduction electrons, the temperature $T_{c}(t)$ and chemical potential $\mu_{c}(t)$ follow from $n_c$ and $U_c$ numerically. Fig.~4b shows the steady-state conduction electron temperature as a function of deposited energy. Neglecting the Auger processes and assuming all energy remains in the conduction band (dashed line), the temperature rises rapidly to extremely large values exceeding 6~eV (70,000~K). Including the interband processes described above (red line) suppresses this rise, saturating near 1~eV. The large density of valence states acts, once accessible, as a heat reservoir of enormous capacity, limiting the peak temperature to around 1.5~eV. Fig.~4c shows the effect of the delay in Auger excitation: initially the conduction electron temperature spikes sharply, but after approximately 120~fs the onset of interband excitation moderates the temperature, pulling it well below the value it would otherwise reach.

We first focus on modelling $\omega_\mathrm{p}$, since through its dependence on both the free-carrier density $n_c$ and the effective mass $m^*$ it directly encodes the competition between the intraband and interband contributions. In Figs.~\ref{fig:model}d--f we compare our model to three traces spanning the full range of pump fluences. At 19~mJ/cm$^{2}$ the Auger contribution is negligible and the intraband-only model reproduces the data, with a smooth monotonic recovery on the electron-phonon timescale. At 57~mJ/cm$^{2}$ the two models diverge: the intraband-only prediction fails to capture the shallow secondary feature and underestimates the recovery, while the full model tracks the data closely. At 131~mJ/cm$^{2}$ the intraband-only model predicts a reduction in $\omega_\mathrm{p}$ of approximately 33\%, whereas the full model reproduces both the depth and asymmetry of the modulation, including the inequality of the two minima. The density of valence states and the Kane band parameters are fixed at 131~mJ/cm$^{2}$ and held constant across all fluences (further fits are shown in the supplementary information). The offset between early and late-time values of $\omega_\mathrm{p}$ visible at all high fluences reflects the residual energy term $U_{\infty}$ and is present in the experimental data of Fig.~\ref{fig:drude_maps}c.

The large transient modulations of $\gamma(t)$ observed in Fig.~\ref{fig:drude_maps}d emerge above the same $\sim$55~mJ/cm$^{2}$ threshold as the interband contribution to $\omega_\mathrm{p}$, a behaviour inconsistent with temperature-dependent scattering models, in which $\gamma$ would be expected to grow smoothly with pump fluence from the lowest fluences. Each Auger excitation event promotes one electron into the conduction band and leaves a hole in the valence band. These holes reside in flat, nearly immobile bands and scatter conduction electrons far more strongly than the equilibrium band structure would suggest. The threshold onset of $\gamma$ mirrors that of the Auger hole population, and we model $\gamma$ using the computed hole density extracted from the $\omega_\mathrm{p}$ fits, scaled by a constant that accounts for the unknown magnitude of hole-mediated scattering. The results are illustrated in Figs.~\ref{fig:model}g--i and the agreement with data is strong, although the model does not capture the details of the dynamics as well as for $\omega_\mathrm{p}$. Notably, the damping will also be impacted by ionised impurity scattering, acoustic phonon scattering, optical phonon scattering  and increased e-e scattering \cite{wangRoleHotElectron2020, bykovTimeDependentUltrafastQuadratic2024}. Here, however, we are primarily interested in capturing the threshold-like evolution of $\gamma$ and its power dependence at high fluences, and in this respect our Auger model works remarkably well. This suggests that interband carrier generation is the dominant driver of damping at high fluences, a conclusion that also supports the success of plasma-frequency-only models at lower fluences where Auger excitation is absent \cite{alamLargeOpticalNonlinearity2016, tiroleSaturableTimeVaryingMirror2022, caspaniEnhancedNonlinearRefractive2016, khurginNonperturbativeNonlinearitiesPerhaps2024}.The unphysical excursion of $\gamma$ below its initial, static value, which is seen at long delays, is an artifact of our simplified uniformly modulated TMM model and is absent when performed with a TMM model that accounts for the depth-dependent pump absorption profile, which returns matching values and never displays the unexpected decrease in $\gamma$. A full comparison between models is presented in the supplementary information.

\section{Discussion}

We have introduced GRUMPY FROG as a route to ultrafast ellipsometry of time-varying films, applying it to ITO near the ENZ condition to simultaneously retrieve the amplitude and phase of the modulated Fresnel coefficients across a decade of pump fluence. Inverting these via a TMM model yields $\omega_\mathrm{p}(t)$ and $\gamma(t)$ directly. Above 55~mJ/cm$^{2}$, both parameters depart sharply from the standard two-temperature model. Notably significant modulations of $\gamma$ are observed, with maxima exceeding 600\% of its equilibrium value, evidence that models neglecting damping dynamics cannot reproduce probe scattering at high fluences. Moreover, the modulation of $\omega_\mathrm{p}$ exhibits a sign reversal, transiently exceeding its equilibrium value by over 3\%. The asymmetry in the $\omega_\mathrm{p}$ minimum constitutes model-independent evidence for an out-of-equilibrium process as any equilibrium description requires two equal minima, in opposition to what we observe. By including an Auger contribution to our two-temperature model, we reproduce the dynamics of $\omega_\mathrm{p}(t)$ and $\gamma(t)$ across the full fluence range, importantly addressing the source of the previously observed anomalous ITO modulation dynamics \cite{choiPathwayOpticalCycleDynamic2025, narimanovUltrafastOpticalModulation2025, unElectronicBasedModelOptical2023}. Finally, the model further predicts a reduction in conduction electron temperature at equilibrium from values implied by two-temperature models, explaining the high damage threshold of sub-wavelength ITO films. 

Additionally, GRUMPY FROG is applicable to any thin-film system producing a significant ultrafast modulation, encompassing epsilon-near-zero media~\cite{alamLargeOpticalNonlinearity2016, yangFemtosecondOpticalPolarization2017}, semiconductor metasurfaces \cite{karlFrequencyConversionTimevariant2020, shcherbakovPhotonAccelerationTunable2019}, plasmonic films \cite{grinblatUltrafastSub30fsAlloptical2019}, phase-change materials~\cite{brahmsDecoupledFewfemtosecondPhase2025}, and two-dimensional systems \cite{bridaUltrafastCollinearScattering2013a, hendryEffectsDueGeneration2025}. This generality is timely given the growing demand for materials and characterisation tools capable of realising the promise of time-varying photonic media, such as time reflection, time diffraction and photonic time crystals~\cite{galiffiPhotonicsTimevaryingMedia2022}.  Past works~\cite{kaipurathOpticallyInducedMetaltodielectric2016, wangRoleHotElectron2020, alamLargeOpticalNonlinearity2016, caspaniEnhancedNonlinearRefractive2016} have relied upon inversion of $R$ and $T$ to study transient material properties, however, GRUMPY FROG enables the characterisation of samples with opaque substrates~\cite{tiroleSaturableTimeVaryingMirror2022}, and without amplifying errors when either $R$ or $T$ signals become small~\cite{caspaniEnhancedNonlinearRefractive2016}. Moreover, GRUMPY FROG intrinsically deconvolves the probe pulse from the data without requiring an independent and reliable pulse measurement or a separate deconvolution step, offering a simplified experiment and analysis procedure.

\section{Conclusion}
In conclusion, by resolving $\omega_\mathrm{p}(t)$ and $\gamma(t)$ about the ENZ wavelength for a sub-wavelength ITO film under high-fluence pumping, we have identified a competition between intraband and interband carrier transitions evident from the asymmetry in the $\omega_\mathrm{p}$ transient. We model this by combining the standard two-temperature model with a non-equilibrium interband mechanism, auger-like excitation, in which hot conduction electrons promote valence electrons into the conduction band. This produces an intensity-tuneable ultrafast decay of the refractive index transient, replacing the slow intraband-only recovery observed at lower fluences. The resulting fluence-gated slit geometry is directly compatible with the requirements of time refraction~\cite{zhouBroadbandFrequencyTranslation2020, bohnSpatiotemporalRefractionLight2021} and time diffraction~\cite{tiroleDoubleslitTimeDiffraction2023, harwoodSpacetimeOpticalDiffraction2025}, and extends naturally to applications requiring well-defined, repeatable refractive index transients, such as photonic time crystals~\cite{lyubarovAmplifiedEmissionLasing2022a, wangExpandingMomentumBandgaps2025, sahaPhotonicTimeCrystals2023} and optical switching~\cite{clericiControllingHybridNonlinearities2017}.


\section*{Funding}
We acknowledge computational resources and support provided by the Imperial College Research Computing Service (http://doi.org/10.14469/hpc/2232). A.~C.~H. acknowledges support from the Val O’Donoghue Scholarship in Natural Sciences. J.~B.~P, S.~H., R.~S., T.~V.~R, J.~S and S.V. acknowledge support from UKRI508 (EP/Y015673). R.~S., S.~Z.~J.~L and S.V. are supported by the European Research Council (ERC) (101201360 - LUMINOUS). J.~W.~G~T. acknowledges support from the Defence Science and Technology Laboratory (DSTL).

\section*{Author Contributions}

\noindent{Conceptualisation}: ACH, JBP, SV \\
Methodology: ACH, JBP, SV, SL, TVR, SARH, YL\\
Software: ACH, JBP, SL, TVR, SARH\\
Investigation: ACH, SL, JS\\
Visualisation: ACH, JBP, RS \\
Writing - Original draft:  ACH, JBP \\
Writing - Reviewing and Editing: All authors contributed \\
Supervision: RS, JBP, SV, JWGT \\

\section*{Disclosures}
The authors declare no conflicts of interest.

\section*{Data Availability}
Data underlying the results presented in this paper may be obtained from the authors upon reasonable request.

\bibliography{TimeDomainBIB}

\end{document}